# Are power-law distributions an equilibrium distribution or a stationary nonequilibrium distribution?

Guo Ran, Du Jiulin

*Department of Physics, School of Science, Tianjin University, Tianjin 300072, China*

**Abstract** We examine whether the principle of detailed balance holds for the power-law distributions generated from the general Langevin equation under the generalized fluctuation-dissipation relation (FDR). With the detailed balance and the generalized FDR, we derive analytically the stationary power-law distribution from the Ito's, Stratonovich's and Zwanzig's Fokker-Planck equations, and conclude that the power-law distributions can either be a stationary nonequilibrium distribution or an equilibrium distribution, which depend on information about the form of the diffusion coefficient function, and the existence and uniqueness of an equilibrium state.

**Key words:** Power-law distributions; detailed balance; generalized fluctuation-dissipation relation; nonequilibrium stationary-state

## 1. Introduction

Power-law distributions have been found frequently in some complex systems, such as the kappa-distributions measured by the observations of the solar wind and space plasmas [1]-[6], and lots of $\alpha$-distributions noted in physics, chemistry and elsewhere like $P(E) \sim E^{-\alpha}$ with an index $\alpha > 0$ [7]-[11]. As the systems displaying the power-law behaviors which cannot be well explained by the traditional statistical mechanics, the investigations about the physical mechanism behind the power-law distributions and their dynamical origins become increasingly attractive, and it is very important for us to understand the nature of many different processes in physical, chemical, biological, technical and their inter-disciplinary fields. Some of the power-law distributions associated with complex systems have been modeled under the framework of nonextensive statistics [12].

In the stochastic dynamical theory of power-law distributions, one needs to analyze the stationary behaviors based on the stochastic differential equations for complex dynamical processes, such as Boltzmann equations, Langevin equations and the associated Fokker-Planck (F-P) equations. Usually, it is difficult to solve a general multi-variable F-P equation for a complex system. Thus the previous works have basically focused on some of single-variable and linear F-P equations [13]-[22]. Most recently, we derived the power-law distributions from a general two-variable Langevin equation, with an inhomogeneous friction and a multiplicative noise, and the associated F-P equations [23, 24], which lead the generalized fluctuation-dissipation relations (FDR) for power-law distributions, a generalized Klein-Kramers equation and a generalized Smoluchowski equation. Based on the relevant statistical theory, one can generalize the transition state theory to the nonequilibrium systems with power-law distributions [22], one can study the mean first passage time for power-law distributions [25], the escape rate for power-law distributions in the overdamped systems [26], and the power-law reaction rate coefficient for an elementary bimolecular reaction [27]. A problem was proposed at the end of Ref.[23]: Do the power-law distributions represent a stationary nonequilibrium distribution or an equilibrium distribution? Further, how do we judge them? If they are a stationary nonequilibrium distribution, do they satisfy the condition of detailed balance? Is the conclusion different between the solutions of the Ito's, Stratonovich's and Zwanzig's F-P equations? In this work, we will study these problems.

The paper is organized as follows. In section 2, we briefly review the power-law distributions generated from the Langevin equation and the associated F-P equations. In section 3, we examine whether those power-law distributions satisfy the condition of detailed balance. In section 4, based on the detailed balance and the generalized FDR, the power-law distributions are obtained by solving the stationary F-P equations. In section 5, we make discussions of the principle of detailed balance, an equilibrium state and a stationary nonequilibrium state. Finally in section 6, the conclusion is presented.

## 2. The power-law distributions from the Langevin equations

The Langevin equations modeling the Brownian particles moving in the inhomogeneous medium with the friction coefficient, $\gamma(x, p)$, and in the potential field, $V(x)$, can be written for the position $x$ and momentum $p$ [23] as

$$\frac{dx}{dt} = \frac{p}{m}, \quad \frac{dp}{dt} = -\frac{dV(x)}{dx} - \gamma(x,p)p + \eta(x,p,t), \tag{1}$$

where $m$ is the particle's mass, and $\eta(x,p,t)$ is multiplicative (space/velocity dependent) noise. As usual, the noise is assumed to be Gaussian and satisfies the zero average and the delta-correlated in time $t$,

$$\langle \eta(x,p,t) \rangle = 0, \quad \langle \eta(x,p,t)\eta(x,p,t') \rangle = 2D(x,p)\delta(t-t'). \tag{2}$$

If $\rho = \rho(x,p,t)$ is the probability distribution function with regard to coordinate $x$, momentum $p$ and time $t$, the associated the Ito's, Stratonovich's and Zwanzig's (or backward Ito's rule [28]) F-P equations can be expressed in a unified form [24] as

$$\frac{\partial \rho}{\partial t} = -\frac{p}{m}\frac{\partial \rho}{\partial x} + \frac{\partial}{\partial p}\left[\frac{dV(x)}{dx} + \gamma(x,p)p + \sigma\frac{\partial D(x,p)}{\partial p}\right]\rho + \frac{\partial}{\partial p}D(x,p)\frac{\partial}{\partial p}\rho, \tag{3}$$

where $\sigma = 1$, $1/2$ and $0$ corresponds respectively to the Ito's F-P equation, the Stratonovich's F-P equation and the Zwanzig's F-P equation. In order to show that the power-law distribution is not caused by selecting different stochastic calculus rules, we deal with the unified form of F-P equations given by Eq.(3) with all the three different stochastic calculus rules. In such a general stochastic dynamics, the power-law distributions can be generated if the friction coefficient $\gamma(x, p)$ and the diffusion coefficient $D(x, p)$ satisfy the generalized FDR [23],

$$D = \gamma m \beta^{-1}(1 - \kappa \beta E), \tag{4}$$

with $\beta=1/kT$, where $T$ is temperature, $E=V(x)+p^2/2m$ is the energy, the parameter $\kappa \neq 0$ measures a distance from equilibrium. The standard FDR, $D=m\gamma\beta^{-1}$, for an equilibrium state can be recovered in the case of $\kappa=0$. The readers can see Ref. [23] for more details about the physical explanations for the generalized FDR.

Nothing has been said about requiring that $\rho(r, p, t)$ must approach an equilibrium distribution for long time if no FDR has been invoked. Based on the generalized FDR, Eq.(4), the power-law stationary solutions of Eq.(3) can be obtained in the following two cases:

(a) For the Zwanzig's F-P equation, the stationary distribution is a function of the energy $E$ and exactly is the power-law $\kappa$-distribution in the form [23],

$$\rho_s(E) = Z_\kappa^{-1}\left(1-\kappa\beta E\right)_+^{\frac{1}{\kappa}}, \tag{5}$$

where $Z_\kappa = \iint dxdp\left(1-\kappa\beta E\right)_+^{1/\kappa}$ is the normalization factor, and $(y)_+=y$ for $y>0$ and zero otherwise.

(b) Generally, if the friction and diffusion coefficients are both a function of the energy $E$, the stationary solution of Eq.(3) is exactly the power-law distribution with two parameters $\kappa$ and $\sigma$ in the form [24]:

$$\rho_s(E) = Z_{\sigma,\kappa}^{-1}\left(1-\kappa\beta E\right)_+^{\frac{1}{\kappa}} D^{-\sigma}, \tag{6}$$

where $Z_{\sigma,\kappa} = \iint dxdp\left(1-\kappa\beta E\right)^{1/\kappa} D^{-\sigma}$ is the normalization factor. Therefore we have also obtained the stationary power-law distribution from the Ito's and Stratonovich's forms F-P equations.

## 3. Examination for the principle of detailed balance

Assuming that a physical system has $n$ degrees of freedom and may be described by a set of variables, $\mathbf{q}=\{q_i\}_{i=1,2,\ldots,n}$; making time reversal, we get another set of variables, $\tilde{\mathbf{q}}=\{\varepsilon_i q_i\}_{i=1,2,\ldots,n}$, where $\varepsilon_i=-1$ or $\varepsilon_i=+1$ depends on whether the sign of variable $q_i$ changes or not under the time reversal transformation. If $\rho(\mathbf{q},t)$ is a probability distribution function of the system and $w(\mathbf{q}',\mathbf{q})$ is a transition probability per second, then the principle of detailed balance [29, 30] generally reads

$$w(\mathbf{q}',\mathbf{q})\rho(\mathbf{q}) = w(\tilde{\mathbf{q}},\tilde{\mathbf{q}}')\rho(\tilde{\mathbf{q}}'). \tag{7}$$

If a physical process can be described by an F-P equation, the principle of detailed balance for that process can also be expressed by using the coefficients in the F-P equation and its stationary solution. A general form of the F-P equation [29] is

$$\frac{\partial}{\partial t}\rho(\mathbf{q},t) = -\sum_i \frac{\partial}{\partial q_i} K_i(\mathbf{q})\rho(\mathbf{q},t) + \frac{1}{2}\sum_{ik}\frac{\partial^2}{\partial q_i \partial q_k} K_{ik}(\mathbf{q})\rho(\mathbf{q},t), \tag{8}$$

where $K_i(\mathbf{q})$ is a drift coefficient and $K_{ik}(\mathbf{q})$ is a diffusion coefficient. We assume that the formal stationary solution of Eq.(8) can be written as

$$\rho_s(\mathbf{q}) = N\exp\left[-\phi(\mathbf{q})\right] \tag{9}$$

with $\phi(\mathbf{q})$, an arbitrary function to be determined. For convenience, hereinafter the above two coefficients are defined respectively by the irreversible drift coefficient,

$$D_i(\mathbf{q}) = \frac{1}{2}\left[K_i(\mathbf{q}) + \varepsilon_i K_i(\tilde{\mathbf{q}})\right], \tag{10}$$

and the reversible drift coefficient,

$$J_i(\mathbf{q}) = \frac{1}{2}\left[K_i(\mathbf{q}) - \varepsilon_i K_i(\tilde{\mathbf{q}})\right]. \tag{11}$$

With Eqs.(7) and (8), it has been proved that the principle of detailed balance holds if and only if there are the following three equations [29, 31, 32]:

$$K_{ik}(\mathbf{q}) = \varepsilon_i \varepsilon_k K_{ik}(\tilde{\mathbf{q}}), \tag{12}$$

$$D_i - \frac{1}{2}\sum_k \frac{\partial K_{ik}}{\partial q_k} = -\frac{1}{2}\sum_k K_{ik}\frac{\partial \phi}{\partial q_k}, \tag{13}$$

$$\sum_i \left(\frac{\partial J_i}{\partial q_i} - J_i \frac{\partial \phi}{\partial q_i}\right) = 0. \tag{14}$$

Also, these three equations can be used to seek for the stationary-state solution [29]. Applying Eq.(8) to Eq.(3), we find the following relations,

$$K_1 = \frac{p}{m}, \quad K_2 = -\left[\frac{d}{dx}V(x) + \gamma p + (\sigma-1)\frac{\partial D}{\partial p}\right], \tag{15}$$

$$K_{22} = 2D, \quad K_{11} = K_{12} = K_{21} = 0, \tag{16}$$

$$D_1 = 0, \quad D_2 = -\left(\gamma p - \frac{\partial D_{even}}{\partial p}\right), \tag{17}$$

$$J_1 = \frac{p}{m}, \quad J_2 = -\left[V'(x) - \frac{\partial D_{odd}}{\partial p}\right], \tag{18}$$

where we have divided the diffusion coefficient into two parts, $D = D_{eve} + D_{oddn}$: one part $D_{even}$ is an even function for $p$, $D_{even}(x,p) = D_{even}(x,-p)$, and the other part $D_{odd}$ is an odd function for $p$, $D_{odd}(x,p) = -D_{odd}(x,-p)$. Theoretically, for an arbitrary diffusion coefficient function we can always do so, and this can be proved straightforwardly. The even diffusion coefficient $D_{even}$ can be a function of even power of $p$, or it can be a function of the kinetic energy. But the odd diffusion coefficient $D_{odd}$ actually does not exist because there is no negative diffusion coefficient.

Now we examine whether the power-law stationary solutions, Eq.(5) and Eq.(6), satisfy the principle of detail balance or not.

For the power-law stationary solution, Eq.(5), in the case (a), according to Eq.(5) and Eq.(9), we obtain the function,

$$\phi = -\ln\left[(1-\kappa\beta E)^{\frac{1}{\kappa}}\right]. \tag{19}$$

Substituting the diffusion coefficients $K_{ik}$ in Eqs.(15) and (16), the irreversible drift coefficients

$D_i$ in Eq.(17), the reversible drift coefficients $J_i$ in Eq.(18), and the function $\phi$ in Eq.(19) into Eqs.(12)-(14), we can determine that the three equations, Eqs.(12)-(14), hold only if the odd diffusion coefficient is $D_{odd}(x,p)=0$, namely, $D(x,p)=D(x,-p)$. In other words, we conclude that the principle of detailed balance holds for the power-law stationary solution, Eq.(5), only if the diffusion coefficient is an even function of $p$.

For the power-law stationary solution, Eq.(6), in the case (b), because the friction and diffusion coefficients are both a function of the energy [24], the diffusion coefficient can be only an even function of $p$, i.e., $D = D_{even}$. According to Eq.(6) and Eq.(9), we get the function,

$$\phi = -\ln\left[(1-\kappa\beta E)^{\frac{1}{\kappa}} D^{-\sigma}\right]. \tag{20}$$

Substituting the diffusion coefficients $K_{ik}$ in Eqs.(15) and (16), the irreversible drift coefficients $D_i$ in Eq.(17), the reversible drift coefficients $J_i$ in Eq.(18), the function $\phi$ in Eq.(20) and the condition $D = D_{even}$ into Eqs.(12)-(14), we can determine that the principle of detailed balance always holds for the power-law stationary solution, Eq.(6).

## 4. The stationary solutions under the detailed balance and the generalized FDR

As we all know, a general F-P equation does have lots of stationary state solutions. Obviously, a question is raised that under the condition of the generalized FDR, Eq.(3), whether Eq.(5) and Eq.(6) are only stationary solutions which satisfy the principle of detailed balance. In order to clarify this question, we solve Eqs.(12)-(14) with the relations, Eqs.(15)-(18). Substituting Eqs.(15)-(18), into Eqs.(12)-(14), the principle of detailed balance turns into these three relations,

$$D(x,p) = D(x,-p), \tag{21}$$

$$\gamma p + \sigma \frac{\partial D}{\partial p} = D \frac{\partial \phi}{\partial p}, \tag{22}$$

$$\frac{p}{m}\frac{\partial \phi}{\partial x} = \frac{dV}{dx}\frac{\partial \phi}{\partial p}. \tag{23}$$

Dividing $D$ on both sides of Eq.(22) and integrating it for $p$, we have

$$\phi = \sigma \ln D - \frac{1}{\kappa}\ln(1-\kappa\beta E) + \frac{1}{\kappa}\ln(1-\kappa\beta V) + f(x), \tag{24}$$

where $f(x)$ is an arbitrary continuously differentiable function of $x$, and we have used the generalized FDR, Eq.(3). Taking Eq.(24) into Eq.(23), we can derive

$$\frac{p}{m}\left[\sigma \frac{\partial \ln D}{\partial x} + \frac{df}{dx} + \frac{\beta dV/dx}{1-\kappa\beta V}\right] = \sigma \frac{dV}{dx}\frac{\partial \ln D}{\partial p}. \tag{25}$$

Here the discussion is made for two cases of $\sigma \neq 0$ and $\sigma = 0$, respectively.

For the case of $\sigma \neq 0$ (i.e., Ito's and Stratonovich's F-P equations), Eq. (25) is a first-order linear partial differential equation, so its general solution has

$$\Phi\left(\sigma \ln D + f(x) + \frac{1}{\kappa}\ln(1-\kappa\beta V), E\right) = 0, \tag{26}$$

where $\Phi$ is an arbitrary two-variable continuously differentiable function, which is an implicit functional relation between the two variables, $\sigma \ln D + f(x) + \frac{1}{\kappa}\ln(1-\kappa\beta V)$ and $E$. Eq. (26) can be written in an explicit relation as

$$\sigma \ln D = g(E) - \left[ f(x) + \frac{1}{\kappa} \ln(1 - \kappa\beta V) \right], \tag{27}$$

where $g(E)$ is an arbitrary single-variable continuously differentiable function. It is shown that only if the diffusion coefficient can be reorganized as the form of Eq. (27), the F-P equations have the stationary solution which obeys the principle of detailed balance. Substituting Eq.(24), Eq.(27) and the generalized FDR, Eq.(3) into Eq.(9), we derived the stationary solution,

$$\rho_s(E) = N \exp(-g(E))(1 - \kappa\beta E)^{\frac{1}{\kappa}}, \tag{28}$$

where $N$ is a normalization factor, and $g(E)$ depends on an exact form of the diffusion coefficient $D$. Or we can rewrite Eq.(28) using $D$ and $f(x)$ instead of $g(E)$ as

$$\rho_s(E) = N \exp(-f(x)) D^{-\sigma} \left( \frac{1 - \kappa\beta E}{1 - \kappa\beta V} \right)^{1/\kappa}, \tag{29}$$

where $f(x)$ is determined also by an exact form of the diffusion coefficient $D$.

For $\sigma = 0$ (i.e., Zwanzig's F-P equation), similarly to the above ways, we can find the power-law stationary solution, which is exactly Eq.(5),

$$\rho_s(E) = N(1 - \kappa\beta E)^{\frac{1}{\kappa}}, \tag{30}$$

with a normalization factor $N$. And note that for this solution the diffusion coefficient $D$ is required to be Eq. (21).

All these results above are discussed under the generalized FDR, Eq.(3), which lead the power-law distributions and the power-law parts in the distribution functions. When we take the limit $\kappa \to 0$, the generalized FDR becomes the standard FDR and the stationary distributions become the traditional exponential form.

## 5. The principle of detailed balance and an equilibrium state

On one hand, in section 4 our results showed that for the F-P equation, if the diffusion coefficient satisfies Eq.(27) for $\sigma \neq 0$ and Eq.(21) for $\sigma = 0$, respectively, i.e., if the diffusion coefficient is an even function of $p$, there exists only one possible stationary solution which satisfies the principle of detailed balance, otherwise there is no stationary solution in the detailed balance. So if an equilibrium exists and is unique, then the power-law distribution, Eq.(29) for $\sigma \neq 0$ or Eq.(30) for $\sigma = 0$, is an equilibrium distribution. On the other hand, in section 3 our results implied that the power-law distribution, Eq.(6), in the case (b) would represente an equilibrium state if the diffusion coefficient was an even function for $p$, $D(x,p)=D(x,-p)$; otherwise if $D(x,p)=-D(x,-p)$, which break the detailed balance, Eq.(6) would be a stationary nonequilibrium distribution. The odd diffusion coefficient $D(x,p)=-D(x,-p)$ implies that it is not a function of the kinetic energy.

As we all know, the principle of detailed balance holds usually only when a system reaches at an equilibrium state. *However, if the system is at a nonequilibrium stationary-state, the detailed balance may either break or hold.* In some experiments it was proved that a nonequilibrium system may approximately obey the principle of detailed balance [33]-[38]. But if the detailed balance is broken, the system must be at a nonequilibrium state. Ref.[39] proved that the broken detailed balance would lead to the entropy production, one of the key signatures as a nonequilibrium stationary-state. Thus, the power-law distributions are not necessarily a stationary nonequilibrium distribution, neither are they necessarily an equilibrium distribution. These two situations are both possible. It is clear that more informations are required about the diffusion coefficient, and the existence and uniqueness of an equilibrium state, if one wants to distinguish the state represented by the power-law distributions,.

## 6. Conclusions

In conclusion, firstly, we have examined whether the power-law distributions, Eq.(5) and Eq.(6), satisfy the principle of detailed balance. They are the stationary solutions of the F-P equations, Eq.(3), under the condition of the generalized FDR, Eq.(4). We showed that the power-law distribution, Eq.(5), in the case (a) satisfy the detailed balance only if the diffusion coefficient is an even function of the momentum $p$, i.e., $D(x,p)=D(x,-p)$, which imply that the diffusion coefficient is a function of the kinetic energy; the power-law distribution, Eq.(6), in the case (b) always satisfy the detailed balance. Additionally, according to FDR (4), the condition $D(x, p)= D(x, -p)$ is equivalently to require $\gamma(x,p)= \gamma(x,-p)$, and then the statement about the even diffusion coefficient for $p$ is also suitable for the even friction coefficient for $p$. Thus the power-law distributions can be a stationary nonequilibrium distribution, and also can be an equilibrium distribution.

Secondly, under conditions of the generalized FDR (4) as well as the principle of detailed balance, we have analytically solved the Ito's, Stratonovich's and Zwanzig's F-P equations, Eq.(3), for $\sigma\neq0$ and $\sigma=0$ respectively. We find that the stationary solution of the F-P equations for $\sigma\neq0$ is the power-law distribution Eq.(28) (or it can be written as Eq.(28)), which is a power-law distribution but has a factor that depends on the diffusion coefficient, and thus it can be regarded as a generalization of the power-law distribution Eq.(6). The stationary solution of the F-P equations for $\sigma=0$ is exactly the power-law distribution Eq.(5). Finally, we have made the discussions about an equilibrium state and the detailed balance.

We conclude that the power-law distribution can be either as a stationary nonequilibrium distribution or as an equilibrium distribution, which depends on the specific form of the diffusion coefficient function, the existence and the uniqueness of an equilibrium. If one wants to distinguish that the power-law distributions represent an equilibrium state or a stationary nonequilibrium state, more informations are required about the forms of diffusion coefficient function, and the existence and uniqueness of an equilibrium state.

**Acknowledgements**

This work is supported by the National Natural Science Foundation of China under Grant No. 11175128 and by the Higher School Specialized Research Fund for Doctoral Program under Grant No. 20110032110058.